\documentclass{elsart}
\begin{document}
\begin{frontmatter}
\title{Quantum chaos in quantum Turing machines}
\author{Ilki~Kim and G\"{u}nter~Mahler}
\address{Institut f\"{u}r Theoretische Physik I, Universit\"{a}t Stuttgart\\
Pfaffenwaldring 57, 70550 Stuttgart, Germany\\
phone: ++49-(0)711 685-5100, FAX: ++49-(0)711 685-4909\\
email: ikim@theo.physik.uni-stuttgart.de}

\begin{abstract}
We investigate a 2-spin quantum Turing architecture, in which discrete local 
rotations $\alpha_m$ of the Turing head spin alternate with quantum controlled 
NOT-opera-\linebreak
tions. We demonstrate that a single chaotic parameter input 
$\alpha_m$ leads to a chaotic dynamics in the entire Hilbert-space.
\end{abstract}
\end{frontmatter}

\section{Introduction}
Chaotic behaviour as an exponential sensitivity to initial conditions in 
a classical non-linear system has attracted a great deal of attention. 
The deterministic chaos, which occurs in 
non-dissipative systems, can typically be found starting from 
regular states as a function of some external control parameter. 
On the other hand, there seems to be no direct analogue to 
chaos in the quantum world, because the Schr\"{o}dinger equation is linear 
in time, and the scalar product between different initial states 
(as a measure of distance) is conserved under unitary evolution. 
Accordingly, the semiclassical 
{\em quantum chaology} \cite{BER85} has been constrained to studying 
some quantum-mechanical ``fingerprints of chaos'' (like spectral properties), 
and non-trivial transitions from the quantum - to classical domain and 
vice versa (e.g., Bohr's correspondence principle). 
Experimental progress 
in mesoscopic physics, e.g. the transport of electrons through so-called 
``chaotic quantum dots'' \cite{KOU97}, has allowed to study a 
quantum-mechanical system in 
a random potential, the results of which give numerical evidence for weak 
chaos (indicated by level repulsion) \cite{SHE94}.\\
\hspace*{1ex}
Recent theoretical and experimental studies in quantum 
information theory and quantum computation~(QC) \cite{STE98} 
should shed new light also on the basic understanding of 
quantum mechanics itself. In~QC one tries to utilize the quantum-mechanical 
superposition and (non-classical) entanglement to 
solve certain classes of problems in a potentially very powerful way. 
While most models of QC have been based on networks of quantum 
gates, which are reminiscent of classical integrated circuits, 
quantum Turing machines~(QTM) \cite{BEN82,DEU85} follow a different line but 
have not shown much potential for future applications up to now. 
In both cases the complexity of 
the computation is characterized by sequences of unitary transformations 
(or the corresponding Hamiltonians~$\hat{H}$ acting during finite time 
interval steps).\\
\hspace*{1ex}
The investigation of quantum chaos based on quantum gate networks 
has so far been proposed e.g. by an implementation of quantum baker's map 
on a 3-qubit NMR quantum computer \cite{SCH98}, or by realizing a 
quantum-mechanical delta-kicked harmonic oscillator 
in an ion trap \cite{GAR97}. In both cases some sort of sensitivity has been 
located with respect to parameters specifying the dynamics 
(e.g., the respective Hamiltonian). 
In this letter we address an iterative map which, though based on standard 
gates, can be thought to be realized\linebreak
as a QTM architecture: 
Local transformations of the Turing head controlled by a Fibonacci-like 
sequence of rotation angles alternate with a quantum-controlled NOT-operation 
with a second spin on the Turing tape. 
This type of control can generate a chaotic quantum propagation 
(Lyapunov exponent, $\ln\frac{1+\sqrt{5}}{2} > 0$) in the 
``classical'' regime \cite{BLU94} which is defined here as the 
Turing head being restricted to an entanglement-free state sequence 
(``primitive'')~\cite{KIM99}. It will be shown that chaos 
in local Bloch-vector space of the Turing head can be found also in the 
quantum-mechanical superposition of those primitives, 
implying entanglement between head and tape as a genuine quantum feature 
(see Fig~\ref{QTM_chaos}). 
Due to this quantum correlation, we can observe a chaotic propagation even 
in the reduced subspace of the Turing tape (``chaos swapping'').

\section{Chaotically driven quantum-Turing machine}
The quantum network~\cite{MAH98} to be considered in detail is composed 
of 2 spins 
$|p\rangle\!^{(\mu)};\,p=-1,1;\,\mu=S,1$ 
(Turing-head $S$, Turing-tape spin 1) so that its 
network-state $|\psi\rangle$ lives in the {4-dimensional} 
Hilbert-space spanned by the product wave-functions 
$|j^{(S)} k^{(1)}\rangle = |jk\rangle$. 
Correspondingly, any (unitary)\linebreak
network-operator can be expanded 
as a sum of product-operators, which may be based on the $SU(2)$-generators, 
Pauli matrices $\hat{\sigma}_j^{(\mu)},\, j=1,2,3$, with $\hat{1}^{(\mu)}$.\\
\hspace*{1ex}
The initial state $|\psi_{0}\rangle$ will be taken to be 
a product of the Turing-head and tape wave-functions. 
For the discretized dynamical description of this externally driven system 
we identify the unitary operators $\hat{U}_{n},\,n=1,2,3,\cdots$ with the 
local unitary transformation on the Turing-head $S$, 
$\hat{U}_{\alpha_m}^{(S)}$, 
and the quantum-controlled-NOT (QCNOT) on ($S,1$), $\hat{U}^{(S,1)}$, 
respectively as follows:
\begin{eqnarray}
&&\hat{U}_{2m-1} = 
\exp{\left(-i \hat{\sigma}_1^{(S)} {\alpha_{m}}/2\right)}\label{us}\\
&&\hat{U}_{2m} = \hat{U}^{(S,1)} = 
\hat P_{-1,-1}^{(S)}\, \hat \sigma_1^{(1)} + 
\hat P_{1,1}^{(S)}\, \hat{1}^{(1)} = 
\left(\hat{U}^{(S,1)}\right)^{\dagger}\,,\label{ub}
\end{eqnarray}
where $\alpha_{m+1} = \alpha_{m} + \alpha_{m-1}$, $\alpha_0 = 0$, and 
$P_{j,j}^{(S)} = |j\rangle\!^{(S)} {}^{(S)}\hspace*{-0.8mm}
\langle j|$ is a (local) projection operator. 
The $m$th Fibonacci number $\alpha_m$ is given by
\begin{equation}
\label{fibonacci}
\alpha_m\, =\, \frac{\alpha_1}{\sqrt{5}} \left(\beta^{m}-\gamma^{m}\right)\,,
\end{equation}
\vspace*{-5mm}
where $\beta := \frac{1+\sqrt{5}}{2},\,\gamma := \frac{1-\sqrt{5}}{2}$. It is 
useful for later calculations to note that 
$\beta^{m+1} = \beta^{m} + \beta^{m-1},\,
\gamma^{m+1} = \gamma^{m} + \gamma^{m-1}$.\\
\hspace*{1ex}
We restrict ourselves to the reduced state-space dynamics of the head~$S$ and 
tape-spin~$1$, respectively, 
\begin{eqnarray}
\sigma_{j}^{(S)}(n)\, &=&\, \mbox{Tr} \left(\hat{\rho}_n^{(S)} 
\hat{\sigma}_j^{(S)}\right)\, =\, 
\langle\psi_{n}|\hat{\sigma}_{j}^{(S)} \otimes \hat{1}^{(1)}
|\psi_{n}\rangle\,,\nonumber\\
\sigma_{k}^{(1)}(n)\, &=&\, \mbox{Tr} \left(\hat{\rho}_n^{(1)} 
\hat{\sigma}_k^{(1)}\right)\, =\, 
\langle\psi_{n}|\hat{1}^{(S)} \otimes \hat{\sigma}_{k}^{(1)}|\psi_{n}
\rangle\,.\label{bloch}
\end{eqnarray}
Due to the entanglement between the head and tape, both will, in 
general, appear to be in a ``mixed-state'', which means that the length of the 
Bloch-vectors in (\ref{bloch}) is less than $1$. 
However, for specific initial states 
$|\psi_0\rangle$ the state of head and tape will remain pure: 
As $|\pm\rangle\!^{(1)} := \frac{1}{\sqrt{2}}\left(|-1\rangle\!^{(1)} \pm 
|1\rangle\!^{(1)}\right)$ are the eigenstates of 
$\hat{\sigma}_{1}^{(1)}$ with $\hat{\sigma}_{1}^{(1)} 
|\pm\rangle\!^{(1)} = \pm |\pm\rangle\!^{(1)}$, 
the QCNOT-operation $\hat{U}^{(S,1)}$ cannot create 
any entanglement, irrespective of the head state 
$|\varphi\rangle\!^{(S)}$, i.e.
\begin{eqnarray}
\label{entangle}
\hat{U}^{(S,1)}\,|\varphi\rangle\!^{(S)} \otimes\,|+\rangle\!^{(1)}\,&=&\,
|\varphi\rangle\!^{(S)} \otimes\,|+\rangle\!^{(1)}\nonumber\\
\hat{U}^{(S,1)}\,|\varphi\rangle\!^{(S)} \otimes\,|-\rangle\!^{(1)}\,&=&\,
\hat{\sigma}_{3}^{(S)} 
|\varphi\rangle\!^{(S)} \otimes\,|-\rangle\!^{(1)}\,.
\end{eqnarray}
As a consequence, for the initial product-states 
$|\psi_0\rangle = |\varphi_0\rangle\!^{(S)} \otimes 
|\pm\rangle\!^{(1)}$\,with 
$|\varphi_0\rangle\!^{(S)} = \exp{
\left(-i \hat{\sigma}_1^{(S)} {\varphi_{0}}/2\right)}\, 
|-1\rangle\!^{(S)}$\, 
the state~$|\psi_n\rangle$ remains a product-state at any step~$n$ 
and the Turing-head then performs a pure-state trajectory (``primitive'') on 
the Bloch-circle $\left(\sigma_{1}^{(S)}(n)=0\right)$
\begin{equation}
\label{product}
|\psi_{n}^{\pm}\rangle = |\varphi_{n}^{\pm}\rangle\!^{(S)} \otimes 
|\pm\rangle\!^{(1)}\,,
\;\;\;\;\left(\sigma_{2}^{(S)}(n)\right)^{2} + 
\left(\sigma_{3}^{(S)}(n)\right)^{2} = 1\,.
\end{equation}
\hspace*{1ex}
It is easy to verify that the Fibonacci relation and the property 
(\ref{entangle}) give for $\sigma_{j}^{(S)}(n)$ 
$\left( \mbox{see eq.}~(\ref{bloch}) \right)$ of 
$|\varphi_{n}^{+}\rangle\!^{(S)} \otimes\,|+\rangle\!^{(1)},\, n = 2m$\,,
\begin{equation}
\label{plus_fibo}
\sigma_{2}^{(S)}(2m|+) = \sin \mathcal{C}_{2m}(+)\,,\;\;\;\;
\sigma_{3}^{(S)}(2m|+) = -\cos \mathcal{C}_{2m}(+)\,,
\end{equation}
where\, $\mathcal{C}_{2m}(+) := {\displaystyle \sum_{j=1}^{m} \alpha_{j}}$\,, 
and for $n = 2m-1$, $\sigma_{k}^{(S)}(2m-1|+) = \sigma_{k}^{(S)}(2m|+)$. 
In order to find the corresponding expression of $\sigma_{k}^{(S)}(n|-)$ for 
$|\varphi_{n}\rangle\!^{(S)} \otimes\,|-\rangle\!^{(1)}$, 
we utilize the 
following recursion relations for the cumulative rotation angle 
$\mathcal{C}_{n}(-)$ up to step $n$
\begin{equation}
\label{recursion_rel}
\mathcal{C}_{2m}(-) = -\mathcal{C}_{2m-1}(-)\,,\;\;\;\;\mathcal{C}_{2m-1}(-) = 
\alpha_{m} + \mathcal{C}_{2m-2}(-)\,.
\end{equation}
Then $\mathcal{C}_{2m}(-), \mathcal{C}_{2m-1}(-)$ are rewritten, 
respectively, as
\vspace*{-5mm}
\begin{eqnarray}
\label{cumulative}
\mathcal{C}_{2m}(-) &=& -\mathcal{C}_{2m-2}(-) - \alpha_{m} = (-1)^{m-1} 
\sum_{j=1}^{m}\,(-1)^{j} \alpha_{j}\nonumber\\
\mathcal{C}_{2m-1}(-) &=& -\mathcal{C}_{2m-3}(-) + \alpha_{m} = (-1)^{m} 
\sum_{j=1}^{m}\,(-1)^{j} \alpha_{j}\,,
\end{eqnarray}
yielding $\sigma_{2}^{(S)}(n|-) = \sin \mathcal{C}_{n}(-),\,
\sigma_{3}^{(S)}(n|-) = -\cos \mathcal{C}_{n}(-)$ 
$\left(\mbox{{\em cf.} (\ref{plus_fibo})}\right)$. 
The Fibonacci property implies that 
$|\varphi_{n}^{-}\rangle\!^{(S)} \otimes\,|-\rangle\!^{(1)}$ is also 
chaotically 
driven as $|\varphi_{n}^{+}\rangle\!^{(S)} \otimes\,|+\rangle\!^{(1)}$ is.\\
\hspace*{1ex}
From any initial state 
$|\psi_{0}\rangle = a^{(+)}|\varphi_{0}^{+}\rangle\!^{(S)}|+\rangle\!^{(1)} + 
a^{(-)}|\varphi_{0}^{-}\rangle\!^{(S)}|-\rangle\!^{(1)}$, 
we then find at step $n$ 
\begin{equation}
|\psi_n\rangle = a^{(+)}|\varphi_{n}^{+}\rangle\!^{(S)} \otimes\,
|+\rangle\!^{(1)}\,+\,a^{(-)}|\varphi_{n}^{-}\rangle\!^{(S)} \otimes\,
|-\rangle\!^{(1)}
\end{equation}
and, observing the orthogonality of the $|\pm\rangle\!^{(1)}$, 
\begin{equation}
\label{super}
\sigma_{k}^{(S)}(n) = |a^{(+)}|^{2}\,\sigma_{k}^{(S)} (n|+)\,+\,
|a^{(-)}|^{2}\,\sigma_{k}^{(S)} (n|-)\,.
\label{lambda_S}
\end{equation}
This trajectory of the Turing-head~$S$ thus appears, for fixed~$n$, as a
decomposition into two Bloch-vectors corresponding to non-orthogonal
pure states, a consequence of the superposition as a quantum feature. 
By using (\ref{cumulative}), (\ref{super}) 
$\left( \mbox{with}\; a^{(+)} = a^{(-)} = 1/\sqrt{2}\right)$ 
we thus obtain for 
$|\psi_0\rangle = |-1\rangle\!^{(S)} \otimes\,|-1\rangle\!^{(1)}$
\begin{eqnarray}
\label{chaotic_driving}
\left(\sigma_{2}^{(S)}(2m),\,\sigma_{3}^{(S)}(2m)\right)\;&=&\;
\cos \mathcal{A}_m \cdot (\sin \mathcal{B}_m ,\,-\cos \mathcal{B}_m)
\nonumber\\
\left(\sigma_{2}^{(S)}(2m-1),\,\sigma_{3}^{(S)}(2m-1)\right)\;&=&\;
\cos \mathcal{B}_m \cdot (\sin \mathcal{A}_m ,\,-\cos \mathcal{A}_m)\,,
\end{eqnarray}
where $\mathcal{A}_m := \alpha_{m}+\alpha_{m-2}+\cdots\,$, $\mathcal{B}_m := 
\alpha_{m-1}+\alpha_{m-3}+\cdots\,$. 
The equation~(\ref{chaotic_driving}) shows that the local 
dynamics of the Turing head is controlled by a ``chaotic'' driving force 
(``input''), because the 
sequences in $\mathcal{A}_m$ and $\mathcal{B}_m$, namely $\{\alpha_{2m}\}$ 
or $\{\alpha_{2m-1}\}$, are in fact both chaotic 
as $\{\alpha_{m}\}$ is. The Bloch-vector 
$\vec{\sigma}^{(S)}(n)$ can alternatively be calculated 
directly from the initial state (here: $|-1,-1\rangle$) and 
for any control angle $\alpha_1$ by using the relations
\begin{eqnarray}
\label{AB}
\mathcal{A}_m &=& \left\{
\begin{array}{ll}
\frac{\alpha_1}{\sqrt{5}}\,\left(\beta^{m+1} - \gamma^{m+1}\right)&
\mbox{\hspace{2.8ex}}m = \mbox{odd}\\
\frac{\alpha_1}{\sqrt{5}}\,\left(\beta^{m+1} - \gamma^{m+1} - \sqrt{5}\right)&
\mbox{\hspace{2.8ex}}m = \mbox{even}
\end{array}
\right.\nonumber\\
\mathcal{B}_m &=& \left\{
\begin{array}{ll}
\frac{\alpha_1}{\sqrt{5}}\,\left(\beta^{m} - \gamma^{m} - \sqrt{5}\right)&
\mbox{\hspace{7ex}}m = \mbox{odd}\\
\frac{\alpha_1}{\sqrt{5}}\,\left(\beta^{m} - \gamma^{m}\right)&
\mbox{\hspace{7ex}}
m = \mbox{even}\,.
\end{array}\right.
\end{eqnarray}

\section{Instability with respect to perturbations}
Now we show that the periodic orbits on the plane 
$\left\{0, \sigma_{2}^{(S)}, 
\sigma_{3}^{(S)}\right\}$ are unstable, which means 
that the dynamics of the Turing head (``output'') is indeed chaotic. 
It is enough to check the periodicity only for step 
$n=2m$: Periodic orbits for $|\psi_0\rangle = |-1\rangle \otimes |-1\rangle$ 
must obey $\mathcal{C}_{2m}(+) = \mathcal{C}_{2m}(-) \stackrel{!}{=} 
2 \pi p,\,p \in \mathbf{Z}$ and $\alpha_{m+1} = \alpha_{1}$ (mod $2 \pi$) 
(one concludes that $\alpha_1$ must be a rational multiple of\, $\pi$). 
By using the Fibonacci numbers (\ref{fibonacci}), we obtain 
$\mathcal{C}_{2m}^{\mbox{per}}(+)$ in (\ref{plus_fibo}) and 
$\mathcal{C}_{2m}^{\mbox{per}}(-)$ in (\ref{cumulative}), respectively, 
for period $= 2m$ as
\begin{eqnarray}
\label{plus_po}
\mathcal{C}_{2m}^{\mbox{per}}(+)\, &=&\, \frac{\alpha_1}{\sqrt{5}}\, 
\left(\beta^{m+2} - \gamma^{m+2} - \sqrt{5}\right)\nonumber\\
\mathcal{C}_{2m}^{\mbox{per}}(-)\, &=&\, \frac{\alpha_1}{\sqrt{5}}\, 
\left(-\beta^{m-1} + \gamma^{m-1} + (-1)^{m} \sqrt{5}\right)\,.
\end{eqnarray}
Now let us consider a small perturbation $\delta$ of the initial 
phase angle $\alpha_0 = 0$, implying 
$|\varphi_0\rangle\!^{(S)} = \exp{\left(-i \hat{\sigma}_1^{(S)} 
{\delta}/2\right)}\, |-1\rangle\!^{(S)}$ 
and a perturbed Fibonacci-like sequence $\{\alpha_m'\}$:
\begin{equation}
\alpha_{0}' = \delta,\, \alpha_{1}' = \alpha_1,\, \alpha_{2}' = \alpha_1 + 
\delta,\, \cdots\,.
\end{equation}
Similarly to (\ref{plus_po}), one finds 
${\mathcal{C}}_{2m}'(\pm) = {\mathcal{C}}_{2m}^{\mbox{per}}(\pm) + 
\Delta {\mathcal{C}}_{2m}(\pm)$, respectively, where 
\begin{eqnarray}
\label{pm_per}
\Delta {\mathcal{C}}_{2m}(+) &=& \frac{\delta}{\sqrt{5}} 
\left(\beta^{m+1} - \gamma^{m+1}\right)\nonumber\\
\Delta {\mathcal{C}}_{2m}(-) &=& -\frac{\delta}{\sqrt{5}} 
\left(\beta^{m-2} - \gamma^{m-2}\right)\,.
\end{eqnarray}
By using (\ref{pm_per}) for $|\psi_0\rangle = |-1\rangle \otimes |-1\rangle$ 
we represent the evolution of the perturbation at the $2m$-th step:
\begin{equation}
\left( \begin{array}{c}
        \Delta \sigma_{2}^{(S)}(2m)\\
        \Delta \sigma_{3}^{(S)}(2m)
        \end{array} \right) = \left( \begin{array}{cc}
                                     M_{11} & 0\\
                                     0 & M_{22}
                                     \end{array} \right) 
                              \left( \begin{array}{c} 
                                    \Delta \sigma_{2}^{(S)}(0)\\
                                    \Delta \sigma_{3}^{(S)}(0)
                                    \end{array} \right)\,,
\end{equation}
where $\Delta \sigma_{2}^{(S)}(0)=\sin \delta,\,\Delta 
\sigma_{3}^{(S)}(0)=-\cos \delta;\,\Delta \sigma_{2}^{(S)}(2m)=
\cos(\delta \alpha_m) \sin$
\linebreak
$(\delta \alpha_{m-1}),\,\Delta \sigma_{3}^{(S)}(2m)=
-\cos(\delta \alpha_m) \cos(\delta \alpha_{m-1});\,M_{11}=
\cos(\delta \alpha_{m}) \sin(\delta \alpha_{m-1})$
\linebreak
$/\sin \delta,\,M_{22}=\cos(\delta \alpha_{m}) 
\cos(\delta \alpha_{m-1})/\cos \delta$, respectively. 
One easily shows
\begin{equation}
\lim_{\delta \to 0} M_{11} = \frac{1}{\sqrt{5}} 
\left(\beta^{m-1} - \gamma^{m-1}\right)
\,,\;\;\;\;\lim_{\delta \to 0} M_{22} = 1\,,
\end{equation}
which means that $M_{11}$ grows exponentially (note that $|\beta| > 1, 
|\gamma| < 1$), 
and the periodic orbit is thus unstable to a small perturbation 
$\delta$ in the external control 
$\left( \mbox{e.g., for period}\,\; n = 40,\, 
\displaystyle{\lim_{\delta \to 0} M_{11} = 4181 \gg 1}, 
\mbox{and see Fig~\ref{stability}}\right)$.\\
\hspace*{1ex}
Strikingly enough, the local dynamics of the Turing tape also shows 
the exponential sensitivity to initial conditions 
$\left(\sigma_1^{(1)}(n) = \sigma_2^{(1)}(n) = 0\right)$:
\begin{equation}
\sigma_{3}^{(1)}(n) = \left\{
\begin{array}{cl}
-\cos\left(\alpha_{\left[\frac{n}{2}\right]+1}\, -\, \alpha_1\, +\, 
\delta_{\left[\frac{n}{2}\right]}^{\mbox{Fib}}\right)&\;\;\;\;n=0,1\;
(\mbox{mod}\; 4)\\
&\\
\cos\left(\alpha_{\left[\frac{n}{2}\right]+1}\, +\, 
\delta_{\left[\frac{n}{2}\right]}^{\mbox{Fib}}\right)&\;\;\;\;n=2,3\;
(\mbox{mod}\; 4)\,,
\end{array}
\right.
\end{equation}
where $\delta_{m}^{\mbox{Fib}} := \frac{\delta}{\sqrt{5}} 
(\beta^m - \gamma^m);\;[a] := n,\,a = n + r,\,n \in \mathbf{Z},\,
0 \leq r < 1$. Similarly to the Turing-head case, it follows for 
$\delta \to 0$ at step $n = 2m+2$, period $2m = 0$ (mod $4$)
\begin{eqnarray}
&\Delta \sigma_{3}^{(1)}(2m+2) = M \cdot \Delta \sigma_{3}^{(1)}(2); 
\lim_{\delta \to 0} M = \frac{1}{\sqrt{5}} \left(\beta^{m+1} - 
\gamma^{m+1}\right)\,\frac{\sin(\alpha_{m+2})}{\sin(\alpha_{1})}&\,,\nonumber
\end{eqnarray}
where $\Delta \sigma_{3}^{(1)}(2) = \cos(\alpha_1 + \delta) - 
\cos(\alpha_{1}),\,
\Delta \sigma_{3}^{(1)}(n) = \cos(\alpha_{m+2} + \delta_{m+1}) - 
\cos(\alpha_{m+2})$, confirming the 
exponential instability of the periodic orbit; it is easily shown that there 
is no periodic orbit with period $2m = 2$ (mod\,$4$). 
Note that the Turing tape can exhibit chaos only by means of the 
entanglement with the head (``chaos swapping''), not as a result of a chaotic 
driving force. The chaotic sequence of Fibonacci-type can be interpreted 
as temporal random (chaotic) ``potential'', in analogy to 
$1$-dimensional ``chaotic quantum dots'' 
in real space \cite{AVI91}. 
It is also interesting to compare this machine 
with a regular QTM \cite{KIM99} which is controlled by a fixed $\alpha$ 
for local transformations of the Turing-head 
by using the Bures metric \cite{HUE92}:
\begin{equation}
D_{\rho \rho'}^{2} := 
\mbox{Tr} \left\{(\hat{\rho} - \hat{\rho}')^2\right\}\,.
\end{equation}
This distance between density matrices, $\hat{\rho}$ and $\hat{\rho}'$, 
lies, independent of the dimension of the Liouville space, between 
0 and 2 [\,see Fig~\ref{2figs}; 
the maximum (squared) distance of 2 applies to pure 
orthogonal states, $D^2 = 2\,(1 - |\langle\psi|\psi'\rangle|^2)$\,]. 
For $\alpha_m = \alpha$ and any $\delta$ the distance 
remains constant; for the Fibonacci-like sequence 
we recognize an initial exponential sensitivity, 
which is eventually constrained, though, by $D^2 \leq 2$.\\
\hspace*{1ex}
The source of the considered chaotic behaviour can be traced back to any
\linebreak 
small perturbation $\delta$ of the initial state $|\psi_0(\delta)\rangle$ 
which 
is directly connected with a perturbed 
unitary evolution, $\hat{U}(\delta)$. This implies that the scalar product 
between different initial states 
(as a measure of distance) is no longer conserved under these evolutions:
\begin{equation}
O' := |\langle \psi_0(\delta)| \hat{U}^{\dagger}(\delta)\,
\hat{U}(0) |\psi_0(0) \rangle|^2\,,\;\;\;\;D^2 = 2 ( 1 - O')\,.
\end{equation}
Thus the initial state is directly correlated to its unitary evolution, which 
can lead to the exponential sensitivity to initial condition, 
whereas there is no chaos in a generic quantum system evolving by a fixed 
$\hat{U}$ even if characterized by chaotic input parameters. 
This $O'$ reminds us immediately of the test function 
$O = |\langle \psi| \hat{V}^{\dagger}(t)\, \hat{U}(t) 
|\psi \rangle|^2$ \cite{PER91}, where $\hat{U}, \hat{V}$ are specified by 
slightly different external parameters (``Peres test''): The corresponding 
parameter-sensitivity has been proposed as a measure to 
distinguish quantum chaos from regular quantum dynamics. 
The origin of chaos in our QTM may thus be alternatively ascribed to a 
perturbed $\hat{V} = \hat{U}(\delta)$ in the control (see also the comment 
by R.~Schack \cite{SCH95}).

\section{Summary}
\begin{figure}
\refstepcounter{figure}\label{stability}
\hspace*{-3.5cm}
\includegraphics{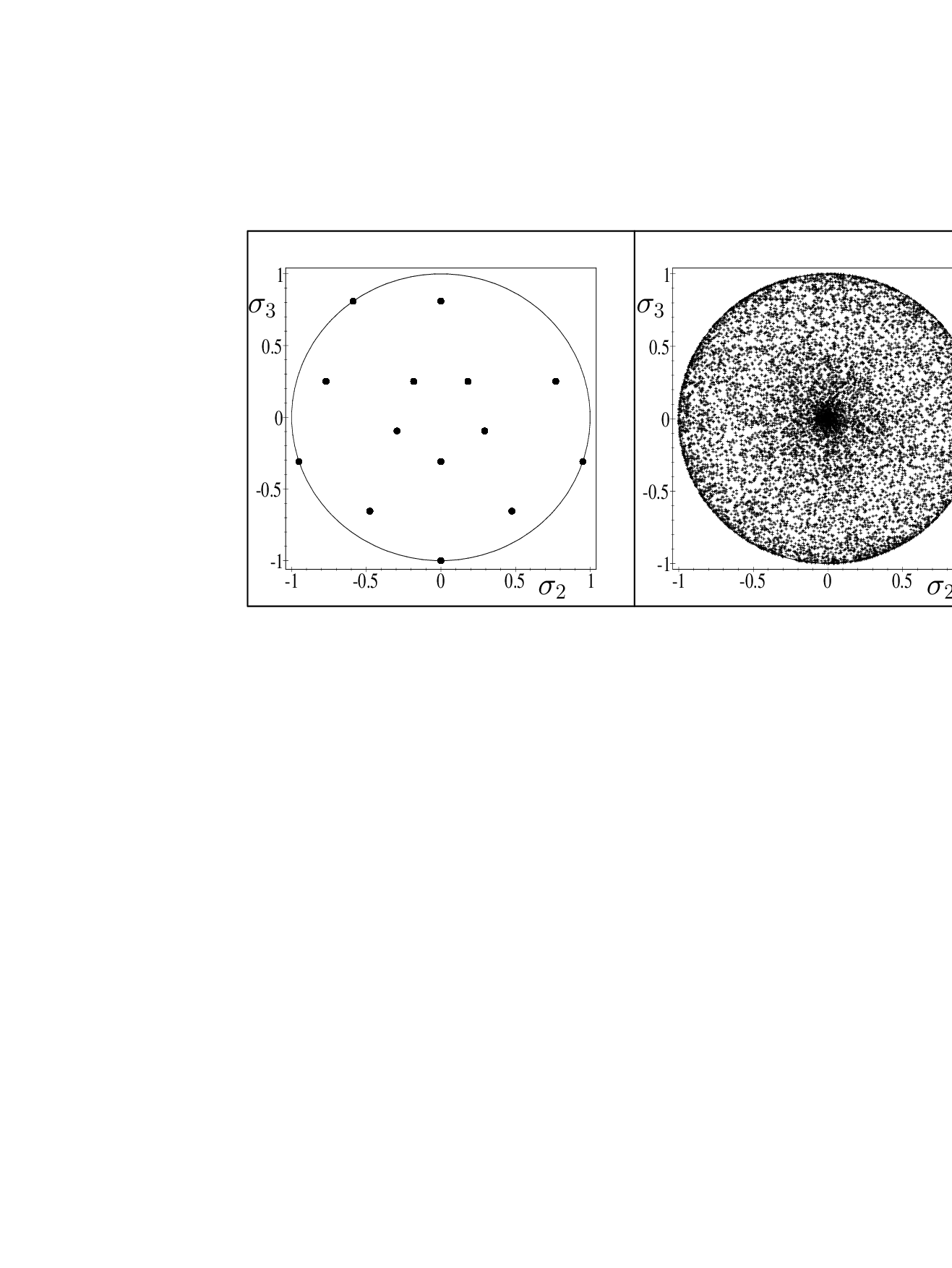}
\end{figure}

In conclusion, we have studied the quantum dynamics of a chaotically 
driven QTM based on a decoherence-free Hamiltonian. 
We have found quantum chaos as a dynamical feature and cumulative loss of 
control in a pure quantum regime. This might be contrasted with the usual 
quantum chaology, which is concerned essentially with 
quantal spectrum analysis of classically chaotic systems 
$\left(\mbox{e.g., level-spacing, spectral rigidity}\right)$. 
As quantum features 
we utilized the superposition principle and the physics of entanglement. 
Our dynamical chaos occurs as a result of 
the superposition and entanglement of a pair of ``classical'' 
(i.e. unentangled) chaotic state-sequences. 
Due to the entanglement, we can observe the 
chaos in any local Bloch-plane. This indicates that patterns in reduced 
Bloch-spheres (a quantum version of a 
Poincar\'{e}-cut, Fig~\ref{stability}) should be useful to 
characterize quantum chaos in a broad class of quantum networks. 
It is worth noting that this kind of control loss is completely different from 
the typical control limit of a quantum network resulting from the 
exponential blow-up of Hilbert-space dimension in which the state evolves 
\cite{FEY82}. It is natural to expect that 
a QTM architecture with an arbitrary number 
of spins on the Turing tape would also exhibit chaos under the same type 
of driving.

\hspace*{1ex}
We would like to thank C.~Granzow, M.~Karremann, A.~Otte and P.~Pangritz 
for stimulating discussions.

\begin{figure}
\refstepcounter{figure}\label{QTM_chaos}
\vspace*{22cm}
\hspace*{-5cm}
\includegraphics{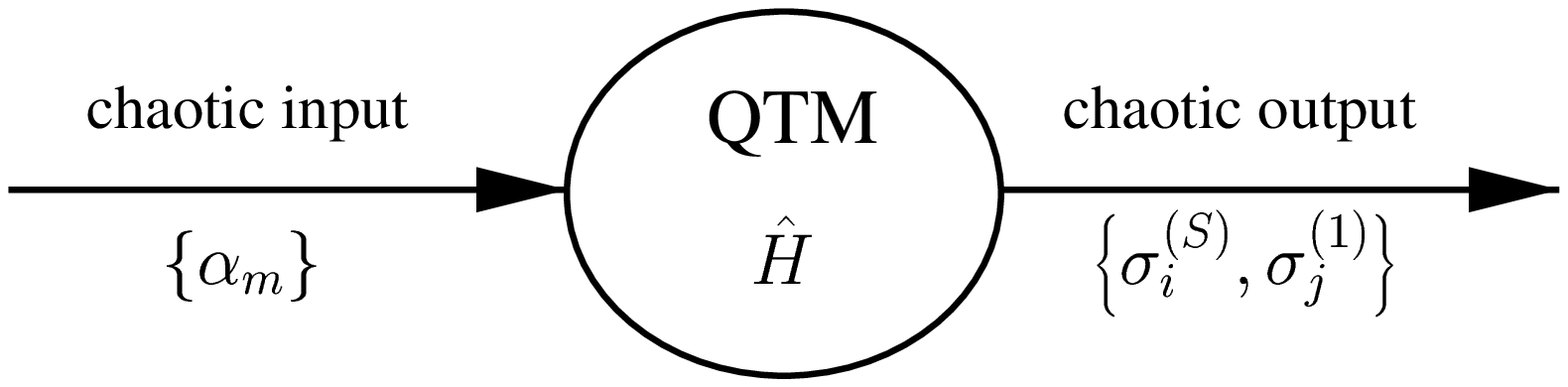}
\end{figure}

Fig.~\ref{QTM_chaos}: Input-output-scheme of our quantum Turing machine~(QTM).

Fig.~\ref{stability}: Turing-head patterns\, 
$\left\{0, \sigma_{2}(n), \sigma_{3}(n)\right\}$\, for initial state 
$|\psi_{0}\rangle = 
|-1\rangle\!^{(S)} \otimes\,|-1\rangle\!^{(1)}$. Left: $\alpha_1 = 
\frac{2}{5} \pi$ (periodic),\, right: $\alpha_1 = 
\frac{2}{5} \times 3.141592654$ (aperiodic) and total step number $n=10000$.

\begin{figure}
\refstepcounter{figure}\label{2figs}
\hspace*{-3.5cm}
\includegraphics{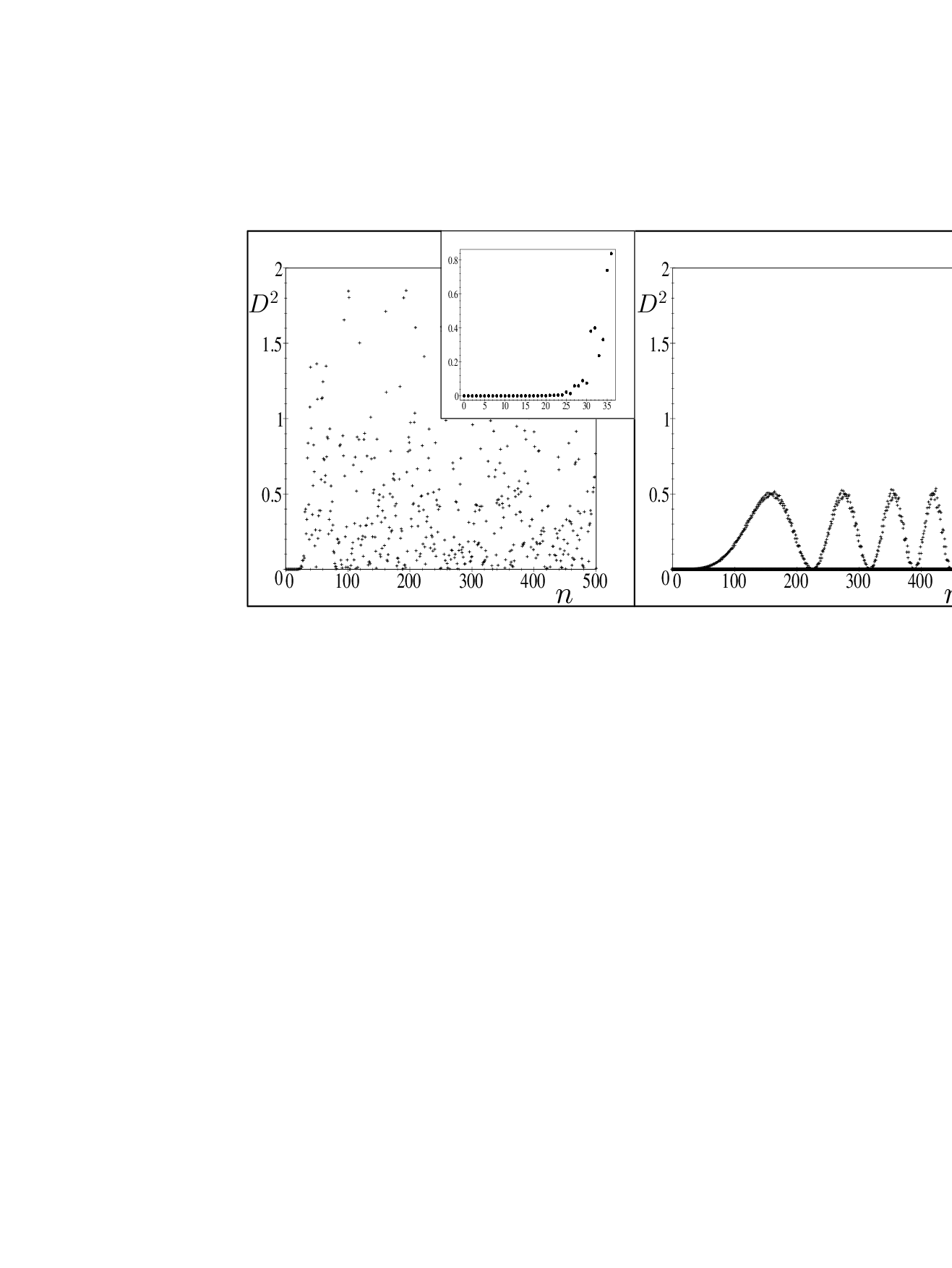}
\end{figure}

Fig.~\ref{2figs}: Evolution of the distance~$D_{\rho \rho'}^2$ 
between Turing-head state with~$\left(\hat{\rho}'\right)$ and 
without~$\left(\hat{\rho}\right)$ perturbation~$\delta$. 
$\alpha_1 = \frac{2}{5} \pi$, 
$|\psi_{0}\rangle = |-1\rangle\!^{(S)} \otimes\,|-1\rangle\!^{(1)}$ for 
$\hat{\rho}$, and $\left(\exp{\left(-i \hat{\sigma}_1^{(S)} 
{\delta}/2\right)}\, |-1\rangle\!^{(S)}\right) \otimes\,|-1\rangle\!^{(1)}$, 
$\delta = 0.001$ for $\hat{\rho}'$. Left: chaotic input according to 
eq.~(\ref{fibonacci}) (inset shows initial behavior in more detail),\, 
right: $\alpha_m = \alpha$ ($D^2 \approx 0$, solid line) and 
$\alpha_{m+1} = 2 \alpha_{m} - \alpha_{m-1}$ (Lyapunov exponent $= 0$) 
(dotted line); the respective distances $D_{\rho \rho'}^2$ for tape-spin~$1$ 
and for total network state~$|\psi_n\rangle$ are similar to those shown.
\end{document}